# Strong Pinning Enhancement in $MgB_2$ Using Very Small $Dy_2O_3$ Additions

S. K. Chen, M. Wei and J. L. MacManus-Driscoll

## Abstract

0.5 – 5.0 wt.% $Dy_2O_3$ was *in-situ* reacted with Mg + B to form pinned $MgB_2$. While $T_c$ remained largely unchanged, $J_c$ was strongly enhanced. The best sample (only 0.5 wt.% $Dy_2O_3$) had a $J_c \sim 6.5 \times 10^5$ $Acm^{-2}$ at 6K, 1T and $3.5 \times 10^5$ $Acm^{-2}$ at 20K, 1T, around a factor of 4 higher compared to the pure sample, and equivalent to hot-pressed or nano-Si added $MgB_2$ at $\leq$ 1T. Even distributions of nano-scale precipitates of $DyB_4$ and MgO were observed within the grains. The room temperature resistivity decreased with $Dy_2O_3$ indicative of improved grain connectivity.

Applications in the temperature window of 20 – 26K are of particular interest for $MgB_2$. A near term goal is MRI coils operating at 2 – 4 T. To date, various mechanical processing methods[1-3] and nano-particle inclusions, in some cases with lattice doping ($Y_2O_3$[4], nano-Si[5] and SiC[6] additions) have been demonstrated to increase the critical current density, $J_c$.

Trace amounts of oxygen during sample preparation can be both detrimental and beneficial to the current carrying properties of $MgB_2$. Oxide nano-phases have great potential to act as pinning centres but only if they are of the correct size and distribution, namely if they are incorporated within the grains and not at the grain boundaries. Formation of MgO and other Mg-B-O phases readily occurs depending on the processing conditions[7-9]. MgO has been shown to lead to isotropic pinning in films deposited in oxygen atmospheres[10] but the adverse effect of excess oxide phases at grain boundaries results in a degradation of grain connectivity[11]. Other structural defects which are effective pinning centres include dislocations[12] and grain boundaries[13].

Previously, rather large additions (10 wt.%) of $Y_2O_3$ (which formed $YB_4$ and MgO nanoparticles when reacted with $MgB_2$) were found to increase both $J_c$ and irreversibility field despite the fact that the samples were nearly insulating indicative of very poor connectivity[4]. The aim of this work is to study much lower levels of rare earth oxide additions in order to enhance intragrain pinning without degrading the grain boundaries. $Dy_2O_3$ was used instead of $Y_2O_3$ since Dy is the cheapest of all the rare earth elements. According to the phase diagram of Dy-B[14], similar dysprosium boride phases as those in the Y-B system[4] form across the composition range.

The starting powders were crystalline magnesium (99.8 %, 325 mesh) from Alfa Aesar, amorphous boron (95 – 97 %) from Fluka, and 1 – 3 µm sized $Dy_2O_3$ (99.99 %) from Reacton. 0.5 - 5.0 wt.% additions of $Dy_2O_3$ were studied. The powders were well mixed by grinding in a mortar for 60 min. Pellets of 5 mm diameter were uniaxially pressed, wrapped with Ta foil in the presence of excess Mg shavings and sintered at 900ºC for 15 min. using heating and cooling rates of 15 ºC/min. X-ray powder diffraction (XRD) patterns in the step-scanning mode $\theta$ - $2\theta$ with 0.05º increment were recorded using a Philips PW1050 diffractometer with a Cu-$K_\alpha$ radiation source. Microstructures were observed using field emission gun scanning electron microscopy (FEG-SEM). High resolution transmission electron microscopy TEM) was undertaken on thinned samples using a JEOL 4000EX MK II microscope operating at 400 kV. Room temperature resistivity was carried out according to the standard four point method. Magnetisation versus temperature measurements were obtained using a commercial Quantum Design DC Magnetic Properties Measurement System (MPMS-XL). Magnetisation hysteresis loops were performed on bar shaped samples with the magnetic field applied parallel to the longest dimension of the sample. Magnetic critical current density was estimated based on the critical state model[15].

Fig. 1 shows XRD patterns of $MgB_2$ with different amounts of $Dy_2O_3$ additions. $MgB_2$ is always the main phase observed. MgO was also observed in all the samples. $DyB_4$ and unreacted $Dy_2O_3$ were found in all the doped samples in increasing amounts with increasing $Dy_2O_3$ additions. Clearly, not all the $Dy_2O_3$ underwent complete reaction with B. Within the limit of calculation error, the *a* and *c* lattice parameters obtained from Rietveld refinements (see Table 1) did not change with $Dy_2O_3$ addition except for the 5 wt.% sample which showed a slight increase in *c*. The geometric density showed that the pellets were slightly more dense with $Dy_2O_3$ addition level (1.44 gcm$^{-3}$ for the 5 wt.% doped sample compared to 1.16 gcm$^{-3}$ for the pure one (Table 1)).

From the FEG-SEM images of the pure and 0.5 wt.% doped samples of Figs. 2a and 2b, a moderate refinement of grain morphology (100 – 200 nm cf. 500 nm) can be seen. However, in the pure samples (Fig. 2a), some fine particles of tens of nm size were also found among the bigger grains.

Fig. 2(c) shows a bright-field TEM image of a 0.5 wt.% Dy$_2$O$_3$ added sample. Nano-sized (~ 10nm) precipitates of DyB$_4$ and MgO were observed inside the MgB$_2$ matrix. A low density of larger precipitates of the size of ~ 70nm were also observed in other regions of the sample. In Fig. 2(c), some precipitates show the so-called 'line of no contrast' which is present for a particle with a spherical strain field when a single reflection is operating. The line of no contrast lies perpendicular to the direction of **g**, as described by Ashby and Brown[16]. Qualitatively, the size and detailed contrast observed depends on the magnitude of the particle radius, constrained strain and the order of the reflection. An increase in any of these parameters leads to an increase in the size of the strain field. Fig 2(d) shows a high resolution image of an intragrain nano inclusion (< 5nm) which can be indexed as DyB$_4$. Both the DyB$_4$ particles *and* associated strain fields can act as effective pinning centres, thus enhancing $J_c(H)$.

Fig. 3 shows the field dependence of the magnetic critical current density measured at 6K. $J_c$ is increased with Dy$_2$O$_3$ additions especially in the low field region, and the best result is achieved for the 0.5 wt.% Dy$_2$O$_3$ additions. $J_c$ was slightly degraded by further additions although it was still higher than the undoped sample. 0.25 wt.% Dy$_2$O$_3$ additions (data not shown) yielded very similar $J_c$'s as for the 5 wt.% doped sample. As shown in Table 1, $J_c$(1T) at 6K and 20K of the 0.5 wt.% sample was increased by a factor of more than 4 compared to the pure sample. As shown in the inset of Fig. 3 and in Table 1, $T_c$ remains largely unchanged. However, the 5 wt.% $Dy_2O_3$ sample showed a slightly reduced $T_c$ of 37.5K. We recall that the *c* parameter for the 5 wt.% $Dy_2O_3$ sample was also marginally higher. These two findings suggest that for the higher doping level there is either a very small level of Dy for Mg substitution, or a perturbation of the B planes by the nanoparticles.

As shown in Table 1, except for the 0.5 wt.% sample, the room temperature resistivity ($\rho$) was reduced with $Dy_2O_3$ additions. This trend was reproduced in a second batch of samples. $\rho$ for the pure sample was higher than very clean bulk $MgB_2$ i.e. 62.2 $\mu\Omega$ cm compared to around 15 $\mu\Omega$ cm[17], However, for the 0.5 wt.% sample $\rho$ was low compared to SiC doped samples, i.e. 75 $\mu\Omega$ cm compared to 522 $\mu\Omega$ cm[17]. At first, the results are surprising since $\rho$ would be expected to increase across the series as a result of increased impurity scattering. An explanation for this is proposed by the reaction equation (1):

$$xDy_2O_3 + Mg + 2B \longrightarrow 2xDyB_4 + (1-4x)MgB_2 + 3xMgO + xMg \quad (1)$$

Equation (1) shows the competing effects of (a) a decrease in $\rho$ because of excess Mg which results because the B is reacted with some Dy and not all the Mg, and (b) an increase in $\rho$ due to extra scattering from the $DyB_4$, MgO and unreacted $Dy_2O_3$. For every five moles of ($DyB_4$ + MgO) formed, there is one mole of Mg produced. The nanoparticles of $DyB_4$ and MgO were found to be largely incorporated *within* the grains. The fact that $\rho$ shows a broadly decreasing trend suggests that the smaller amounts of evolved Mg were incorporated in the grain boundary regions. The amount of unreacted Mg is sufficiently small so as not be detectable by X-ray diffraction. The presence of unreacted Mg in MgB$_2$ has previously been shown to reduce $\rho$ greatly[18].

Fig. 4 compares the field dependence of the magnetic $J_c$ from this work to MgB$_2$ samples which have shown improved low-field $J_c$'s, namely hot isostatically pressed (HIPed), mechanically milled (nano-grain) material, either pure[1] or carbon doped[2]; HIPed pure (micro-grain) material[5]; and cold uniaxially pressed nano-Si[5] and SiC doped[6] samples. With the exception of the pure HIPed sample[5], for all the literature samples in Fig. 4 the relative $J_c$ enhancement from either pinning or doping is uncertain, i.e. samples containing Si, and/or C, and ball milled samples all show slightly reduced $T_c$'s indicative of doping, and this explains the good $J_c$ performance at > 2T. It is the < 1-2T regime that needs to be studied carefully to assess the influence of the pinning alone.

At 6K, 0-2T, the $J_c$ for the 0.5 wt.% Dy$_2$O$_3$ added sample is slightly higher than the HIPed pure sample[5]. At self-field, $J_c$ cannot be assessed accurately from magnetic data because of the flux jumping and also the sample is not fully field penetrated. However, at 20K and below 2T, the $J_c$ of the Dy$_2$O$_3$ sample outperforms the ball milled nano-grain pure[1] and carbon doped[2] samples. It is similar to the pure HIP'ed[5] and nano-Si doped[5] samples at ≤ 1T. The best sample from this work together with the 5 wt.% nano-Si doped[5] sample likely represent the most highly pinned MgB$_2$ bulk material. By hipping such pinned materials, it should be possible to achieve $J_c$'s in excess of 1 MAcm$^{-2}$ at 20K, 2T. Alternately, through C doping of the pinned material, there is the possibility to achieve $J_c$'s of $10^5$ Acm$^{-2}$ at ~ 4T, 20K.

In summary, a very simple method of reacting very small amounts of Dy$_2$O$_3$ (~ 0.5 wt. %) with Mg and B to significantly increase pinning in MgB$_2$ has been demonstrated. $T_c$ is preserved at around 38K. High resolution TEM imaging shows nano-scale precipitates of $DyB_4$ and MgO within the grains. The surprising decrease in room temperature resistivity with $Dy_2O_3$ additions is believed to be due to unreacted Mg which decreases the intergrain resistivity.

We are grateful to EPSRC, UK for funding this work. S. K. Chen acknowledges UPM for the financial assistance.

Table 1    Lattice constants, density, room temperature resistivity, $T_c$ and $J_c$ of the pure and $Dy_2O_3$ added samples.

| $Dy_2O_3$ addition (wt.%) | Lattice parameters | | Density (gcm$^{-3}$) | $\rho_{290K}$ ($\mu\Omega$ cm) | $T_c$ (K) | $J_c$ (Acm$^{-2}$) | |
|---|---|---|---|---|---|---|---|
| | $a$(Å) | $c$(Å) | | | | 6K, 1T | 20K, 1T |
| 0 | 3.0846(4) | 3.5253(3) | 1.16 | 62.2 | 38.0 | 1.5 x 10$^5$ | 8.0 x 10$^4$ |
| 0.5 | 3.0839(6) | 3.5254(6) | 1.27 | 75.0 | 38.0 | 6.5 x 10$^5$ | 3.5 x 10$^5$ |
| 1.0 | 3.0845(4) | 3.5254(4) | 1.27 | 46.5 | 38.0 | 4.9 x 10$^5$ | 2.5 x 10$^5$ |
| 2.0 | 3.0843(4) | 3.5254(4) | 1.32 | 42.6 | 38.0 | 5.0 x 10$^5$ | 2.3 x 10$^5$ |
| 5.0 | 3.084(1) | 3.526(1) | 1.44 | 24.0 | 37.5 | 4.2 x 10$^5$ | 2.2 x 10$^5$ |

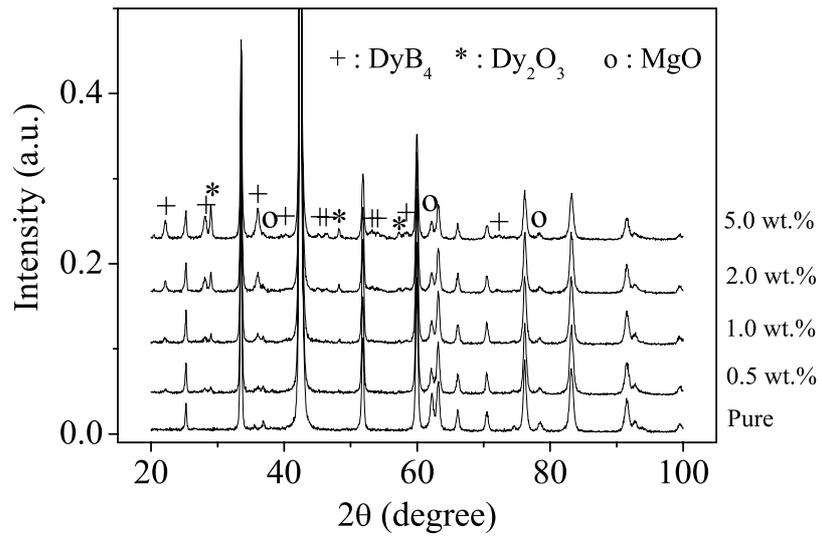

Fig. 1 X-ray diffraction patterns of the pure and $Dy_2O_3$ added samples.

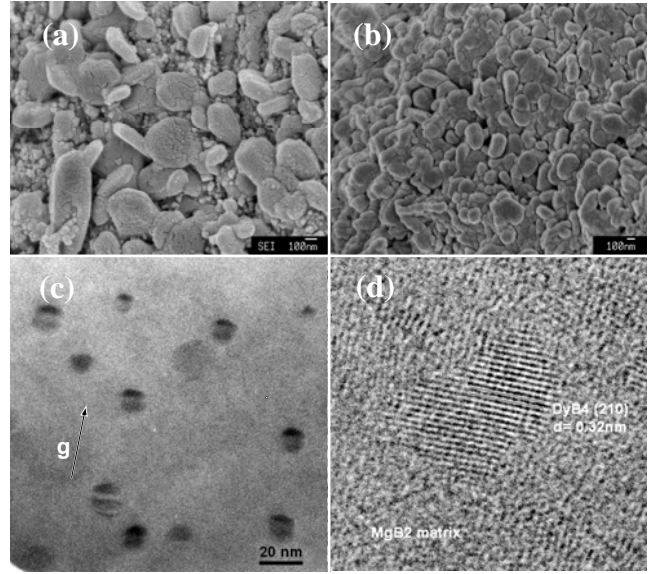

Fig. 2 FEG-SEM images of the (a) pure $MgB_2$ and (b) sample added with 0.5 wt.% of $Dy_2O_3$. (c) Bright-field TEM micrograph of $MgB_2$ matrix strain field contrast with a high order matrix reflection. The reciprocal lattice vector **g** is indicated by an arrow. (d) HRTEM image showing the (210) lattice fringe of a nano-sized $DyB_4$ precipitate.

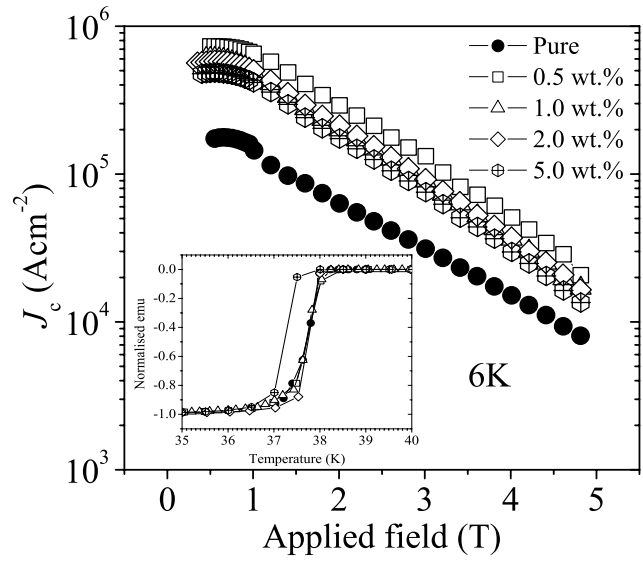

Fig. 3 Field dependence of magnetic $J_c$ at 6K. Inset: Temperature dependence of normalised magnetic moment.

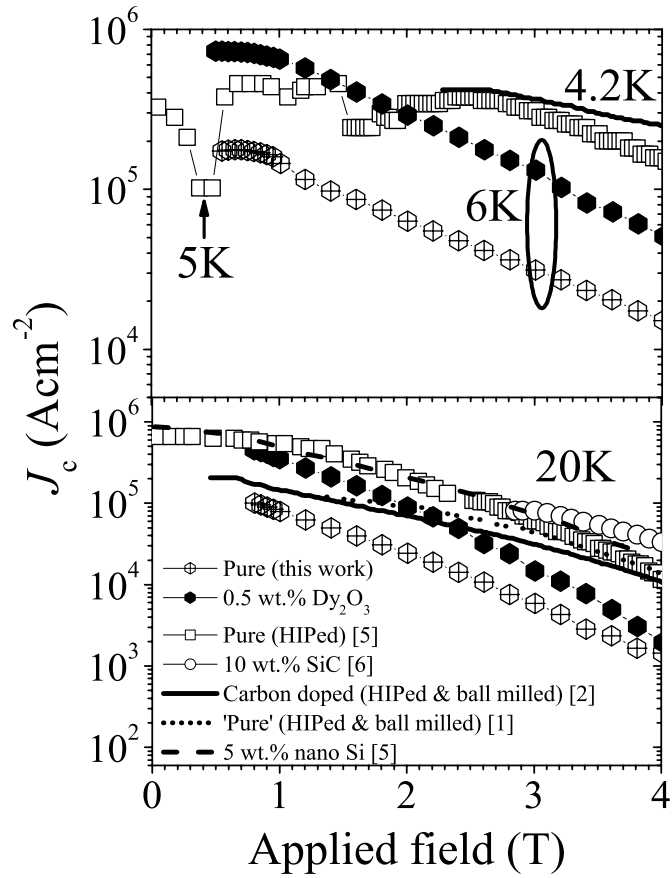

Fig. 4  $J_c$ of the pure and best sample in comparison with those from references 1, 2, 5 and 6. Data on $J_c$ below 6K are not available in some of the cited references.